\documentclass[aps,pra,reprint,superscriptaddress,showpacs,floatfix,preprintnumbers,showkeys]{revtex4-1}

\usepackage{graphicx}
\usepackage{color}
\usepackage{amssymb}
\usepackage{amsmath}
\usepackage{epsfig}
\usepackage{bm}
\usepackage[american]{babel}
\usepackage[colorlinks,breaklinks=true,urlcolor=blue,citecolor=blue,linkcolor=blue,pdfstartview=FitH,pdfpagemode=UseNone]{hyperref}
\DeclareGraphicsExtensions{.pdf,.png,.jpg}
\DeclareMathOperator*{\sumsum}{\sum\sum}

\newcommand{\phase}{\varphi}
\newcommand{\ket}[1]{\left|#1\right\rangle}
\newcommand{\avrtopind}{\bar{\mathcal{Q}}}
\newcommand{\topind}{\mathcal{Q}}
\newcommand{\ntot}{n_{\mathrm{tot}}}
\newcommand{\rabidensity}{\varepsilon_{\mathrm{R}}}

\begin{document}

\title{Skyrmionic vortex lattices in coherently coupled three-component Bose--Einstein condensates}
\date{August 12, 2016}
\author{Natalia~V.~Orlova}
\affiliation{Departement Fysica, Universiteit Antwerpen, Groenenborgerlaan 171, B-2020 Antwerpen, Belgium}
\author{Pekko~Kuopanportti}
\affiliation{School of Physics and Astronomy, Monash University, Victoria 3800, Australia}
\author{Milorad~V.~Milo\v{s}evi\'{c}}\email{milorad.milosevic@uantwerpen.be}
\affiliation{Departement Fysica, Universiteit Antwerpen, Groenenborgerlaan 171, B-2020 Antwerpen, Belgium}

\begin{abstract}
We show numerically that a harmonically trapped and coherently Rabi-coupled three-component Bose--Einstein condensate can host unconventional vortex lattices in its rotating ground state. The discovered lattices incorporate square and zig-zag patterns, vortex dimers and chains, and doubly quantized vortices, and they can be quantitatively classified in terms of a skyrmionic topological index, which takes into account the multicomponent nature of the system. The exotic ground-state lattices arise due to the intricate interplay of the repulsive density--density interactions and the Rabi couplings as well as the ubiquitous phase frustration between the components. In the frustrated state, domain walls in the relative phases can persist between some components even at strong Rabi coupling, while vanishing between others. Consequently, in this limit the three-component condensate effectively approaches a two-component condensate with only density--density interactions. At intermediate Rabi coupling strengths, however, we face unique vortex physics that occurs neither in the two-component counterpart nor in the purely density--density-coupled three-component system.
\end{abstract}
\preprint{DOI: \href{http://dx.doi.org/10.1103/PhysRevA.94.023617}{10.1103/PhysRevA.94.023617}}
\maketitle

\section{Introduction}
\label{sc:introduction}

Since the experimental creation of large vortex lattices in rotating single-component Bose--Einstein condensates~(BECs) of atomic gases~\cite{Abo-Shaeer20042001,PhysRevLett.89.100403,Zwierlein.nature03858}, there has been growing interest in exploring the rotational response of BECs with multiple components~\cite{[{For a review, see }]Kas2005.IJMPB19.1835}. The physics of such superfluid mixtures is more involved because the competing effects enter not only via the self-interaction of a single component but also through intercomponent interactions~\cite{pethick,RevModPhys.81.647}. In terms of finding the energy-minimizing vortex configuration, this means that the relative positioning of vortices in different components also profoundly affects the total energy. In addition, two different regimes can be distinguished: the immiscible regime, where the components segregate into nonoverlapping phases, and the miscible regime with interpenetrating BECs~\cite{RevModPhys.81.647}. 

The versatility of the emerging ground-state vortex structures is apparent already in the simplest multicomponent case, the two-component BEC, which has been realized experimentally, e.g., with two hyperfine spin states of atoms of the same species~\cite{PhysRevLett.81.1543,PhysRevLett.83.2498,PhysRevLett.83.3358,PhysRevA.59.R31,PhysRevLett.85.2413,PhysRevLett.87.080402,PhysRevLett.106.065302,PhysRevLett.111.264101}. A rapidly rotating miscible two-component BEC with equally populated and repulsively interacting components was shown to form vortex lattices whose geometry can change from the usual triangular (or, as it is also called, hexagonal) to square, with the lattice unit cells of the two components displaced relative to each other~\cite{PhysRevLett.88.180403,PhysRevLett.91.150406}. Subsequently, a two-component mass-imbalanced BEC with attractive intercomponent interactions was shown to host vortex lattices that vary from a square lattice to a triangular lattice of vortex pairs (dimers)~\cite{PhysRevA.85.043613}. In the immiscible regime corresponding to strong intercomponent repulsion, rotating harmonically trapped two-component BECs undergo phase separation, which can lead to serpentine vortex sheets~\cite{PhysRevA.79.023606} or, when the components are unequally populated, to a giant vortex surrounded by a ring of single-quantum vortices~\cite{PhysRevA.77.033621}. The ground states of rotating two-component BECs can also host spin-texture skyrmions~\cite{Alkhawaja.nature411,PhysRevA.77.033621,PhysRevA.84.033611,PhysRevA.91.043605} or solitary multiquantum vortices~\cite{PhysRevA.91.043605}.

Moreover, a two-component BEC consisting of two hyperfine spin states of the same atom can be coupled not only by density--density interactions but also coherently, so that the phases of the complex-valued condensate wave functions are no longer independent of each other. When rotated, such Rabi-coupled  two-component BECs were found to contain vortex dimers (which can also be interpreted as meron pairs~\cite{PhysRevLett.93.250406,PhysRevA.71.043611}) and multidimer bound states composed of four or six individual vortices from different components~\cite{PhysRevLett.111.170401}. The dimer and multidimer bound states were shown to emerge due to the Rabi coupling giving rise to energy-costing domain walls in the relative phase between the two components~\cite{PhysRevA.65.063621,PhysRevLett.111.170401}. In effect, the domain walls confine vortices in different components into bound vortex molecules~\cite{PhysRevA.65.063621,PhysRevLett.111.170401}. With increasing strength of the Rabi coupling, the domain wall between the bound vortices shrinks and eventually vanishes, merging the constituent vortices into an ``integer vortex" (with the same, coincident phase winding in each component)~\cite{PhysRevA.65.063621,PhysRevLett.107.197001,PhysRevLett.111.170401}. Thus, the Rabi coupling induces an attractive interaction between same-sign vortices in different components. For detailed studies of vortex--vortex interactions in multicomponent BECs, we refer to the works of Eto \emph{et al.}~\cite{Eto2011.PRA83.063603} and Dantas~\emph{et al.}~\cite{PhysRevA.91.023630}.

The sign of the Rabi coupling does not play a role in the ground state of the two-component system, since the Rabi energy can always be reversed in sign, while leaving all other energies unchanged, by multiplying either one of the two condensate wave functions by $-1$. This implies that the overall Rabi coupling in the two-component ground state will always be nonrepulsive. However, the situation becomes significantly more involved when there are more than two coherently coupled components in the system. For instance, in the case of three Rabi-coupled BEC components, there exist parameter regimes where it is impossible for all three pairwise Rabi energies to be maximally attractive at the same point in space, leading to intrinsic phase frustration. The subtle interplay between the various interactions suggests that when such a frustrated system is set in rapid rotation, highly unconventional vortex lattices may appear in the rotating ground state. So far, studies of vortices in Rabi-coupled three-component BECs have focused on states with only a few vortices and have demonstrated, for example, the existence of stable vortex trimers~\cite{PhysRevA.91.023630,PhysRevA.85.053645,Eto2013.EPL103.60006,Nit2014.JLTP175.177}. The possible structures of ground-state vortex lattices in these systems, however, have remained unexplored.

In this work, we investigate the rotational response of coherently coupled three-component BECs in the parameter regime where the Rabi energies necessarily exhibit intrinsic phase frustration. Specifically, we show that the interplay of the intrinsic Rabi frustration with the other interactions and superfluidity of the BEC can result in the emergence of exotic vortex lattices in the rotating ground state of the harmonically trapped system. Fixing the relative phases of two pairs already fixes the relative phase of the remaining pair. This can result in the suppression of some of the three pairwise Rabi couplings and generally leads to the existence of phase-frustrated vortex lattices with unconventional features such as zig-zag patterns, vortex chains, and doubly quantized vortices. In the limit of strong Rabi coupling, the phase frustration causes the  three-component BEC to behave effectively as a two-component BEC with only density--density interactions. Consequently, we also observe a hexagonal-to-rectangular transition in the ground-state vortex lattice in agreement with previous results for the two-component system~\cite{PhysRevLett.88.180403,PhysRevLett.91.150406,PhysRevA.84.033611}. 

The overall repulsive intercomponent interaction tends to favor interlacing of vortices in different components, splitting integer composite vortices into separate entities. Although each component still has the same vortex density determined by the external rotation frequency, the interlaced vortex lattices of the multicomponent system acquire a skyrmionic character and therefore can no longer be satisfactorily described by mere vortex winding numbers. For this reason, we invoke a skyrmionic topological index defined in terms of a $\mathbb{C} P^2$ invariant~\cite{PhysRevB.87.014507} and use it to classify the observed nontrivial states. This classification has a broader scope in multicomponent quantum physics, since the relative phase frustration and vortex interlacing appear not only in multicomponent BECs but also in multiband superconductors~\cite{0295-5075-87-1-17003,PhysRevB.81.134522,PhysRevB.87.134510,PhysRevB.85.064516,Yanagisawa2011675}, where they are associated with, e.g., fractional vortices~\cite{ProgTheorPhys102.965}, solitons~\cite{PhysRevLett.107.197001}, skyrmions~\cite{PhysRevB.87.014507,PhysRevB.90.064509,PhysRevB.89.024512}, and vortex sheets~\cite{PhysRevLett.92.157001}.  

The remainder of the article is organized as follows. In Sec.~\ref{sc:theory}, we outline the theoretical description of the coherently coupled rotating three-component BECs. Section~\ref{sc:results} presents our main results, summarized in two phase diagrams of skyrmionic vortex lattices, with the discovered domains illustrated by representative examples of ground-state solutions. The physical interpretation of the obtained lattices is given in terms of the topological index~(Sec.~\ref{subsc:phase_diagrams}), intrinsic phase suppression of the Rabi coupling~(Sec.~\ref{subsc:phase_suppression}), and domain walls~(Sec.~\ref{subsc:domain_walls}). Finally, we summarize our findings and discuss the outlook in Sec.~\ref{sc:conclusions}.

\section{Theoretical framework}
\label{sc:theory}

Our starting point is a harmonically trapped three-component BEC consisting of three different spin states of a single atomic species, coupled coherently to each other. We use the standard zero-temperature mean-field approach~\cite{pethick} and describe the condensate with three complex-valued wave functions $\Psi_i$, where $i\in\left\{1,2,3\right\}$. For simplicity, we focus on the case of a highly oblate cylindrically symmetric trapping potential with the harmonic trap frequencies satisfying $\omega_{z}^{(i)} \gg \omega_{x}^{(i)}=\omega_{y}^{(i)} \equiv \omega$, which implies that  the BEC is quasi-two-dimensional and the $z$ dependence can be integrated out. Assuming that the system is set into rotation about the $z$ axis with angular frequency $\Omega$, we write the two-dimensional Gross-Pitaevskii (GP) energy functional~\cite{PhysRevA.85.053645} in the rotating frame of reference as
\begin{equation}
\begin{aligned}\label{eq:func_en}
E &=  \int\Bigg[ \sum_{i=1}^3 \Big(  \frac{\hbar^2}{2m}|\nabla \Psi_i|^2+\frac{1}{2}m\omega^2 r^2 |\Psi_i|^2
 \\&- \Omega\Psi^{\ast}_i L_z \Psi_i\Big) +\frac{1}{2}\sum_{i=1}^3 \sum_{j=1}^3 g_{ij} |\Psi_i|^2 |\Psi_j|^2  
 \\&-\hbar \sumsum_{i\neq j} \omega_{ij}\Psi^{\ast}_i\Psi_j \Bigg] d^2 r, 
\end{aligned}
\end{equation}
where $m$ is the mass of the atoms, $r^2=x^2+y^2$, and $L_z=-i\hbar(y\partial_x -x \partial_y)$ is the angular momentum operator. The local density--density interactions are characterized by the intracomponent and intercomponent coupling constants $g_{ii}$ and $g_{ij}\,(=g_{ji},\ i\neq j)$, respectively. We assume that $g_{11}=g_{22}=g_{33}$ and $g_{12}=g_{13}=g_{23}$. The additional coupling constants $\omega_{ij}\, (=\omega_{ji}\in\mathbb{R})$ in Eq.~\eqref{eq:func_en} interlink the phase angles of the three wave functions and are referred to in the literature as the effective Rabi frequencies~\cite{PhysRevLett.93.250406,PhysRevA.71.043611,PhysRevLett.111.170401,PhysRevA.65.063621,allen,PhysRevA.52.2155}. Accordingly,  we call the last term in Eq.~\eqref{eq:func_en} the Rabi energy and denote the corresponding energy density as $\rabidensity=\sumsum_{i<j} \varepsilon_{ij}$, where $\varepsilon_{ij}= -\hbar \omega_{ij}(\Psi^{\ast}_i\Psi_j + \Psi^{\ast}_j\Psi_i)$ are the pairwise Rabi energy densities. The  Rabi term describes a coherent coupling induced by an external driving field, which allows atoms to change their internal state coherently, and has been achieved experimentally for two-component BECs by means of two-photon transitions as reported, e.g., in Refs.~\cite{PhysRevLett.81.1543,PhysRevLett.83.2498,PhysRevLett.83.3358}.

Variation of Eq.~\eqref{eq:func_en} with respect  to each $\Psi^{\ast}_i$ leads  to three coupled time-independent GP equations:
\begin{equation}
 \begin{aligned}
 & \bigg(-\frac{\hbar^2}{2m} \nabla^2 + \frac{1}{2} m\omega^2 r^2 -\Omega L_z  -\mu  \\ & + \sum_{j=1}^3 g_{ij}|\Psi_j|^2\bigg)\Psi_i-\hbar \sum_{j(\neq i)}\omega_{ij}\Psi_{j}=0,
 \label{eq:GP}
 \end{aligned}
\end{equation}
where $i\in \{1,2,3\}$. Here $\mu$ is a chemical potential enforcing the constraint  
\begin{equation}
\int \sum_{i=1}^3 |\Psi_i({\bf r})|^2 d^2 r = N, \label{eq:norm}
\end{equation}
since we consider a coherently coupled system whose Hamiltonian conserves the total particle number $N=\sum_i N_i$ but not the componentwise numbers $N_i  = \int|\Psi_i|^2 d^2 r$.
 
In order to obtain dimensionless quantities for the numerics, we measure length in units of the radial harmonic oscillator length $a_r=\sqrt{\hbar/m \omega}$ and energy in units of $\hbar\omega$. We parametrize the interactions by the two dimensionless quantities $g=g_{11} m N/3 \hbar^2$ and $\sigma=g_{12}/g_{11}$ and consider only the repulsively interacting miscible system with $0<\sigma\leq 1$. Then dimensionless GP equations take the form
\begin{equation}
 \begin{aligned}
 \bigg(&-\frac{1}{2} \tilde{\nabla}^2 + \frac{1}{2}\tilde{r}^2 + g|\tilde{\Psi}_i|^2-\frac{\Omega}{\omega} \tilde{L}_z -\frac{\mu}{\hbar\omega} 
 \\ &+ \sigma g \sum_{j(\neq i)}|\tilde{\Psi}_j|^2\bigg)\tilde{\Psi}_i   - \sum_{j(\neq i)}\frac{\omega_{ij}}{\omega}\tilde{\Psi}_{j}=0,
 \label{eq:GP_cast}
\end{aligned}
\end{equation}
where $\tilde{\Psi}_i=N^{-1/2} \sqrt{3} a_r \Psi_i$ and $\tilde{r}=r/a_r$. Our numerical analysis of the three-component BEC is based on solving Eqs.~\eqref{eq:GP_cast} with link-variable discretization~\cite{PhysRevA.78.053610} and a relaxation method.

\section{Numerical results}
\label{sc:results}

We have numerically solved the GP equations of the rotating three-component BEC in the presence of both density--density and Rabi couplings~[Eqs.~\eqref{eq:GP_cast}]. In all the states we present, we have fixed the intracomponent coupling strength to $g=g_{11} m N/3 \hbar^2=2115$ and the rotation frequency to $\Omega=0.97 \omega$. 
On the other hand, we treat the relative intercomponent density--density coupling strength $\sigma=g_{12}/g_{11}$ and the Rabi frequency $\omega_{12}$ as tunable parameters in order to study their effect on the ground-state vortex lattices of the system. We limit our analysis to repulsive intercomponent interactions in the miscible regime, $0 < \sigma \leq 1$, and take $\omega_{12} < 0$. We assume the other two Rabi frequencies $\omega_{13}$ and $\omega_{23}$ to be equal and consider two different fixed positive values, $\omega_{13}=\omega_{23}=0.01\omega$ and $\omega_{13}=\omega_{23}=0.05\omega$, since these already convey many of the key phenomena associated with the phase-frustrated Rabi coupling. Hence, we end up with two different fixed parameter sets $\left(g,\Omega/\omega,\omega_{13}/\omega,\omega_{23}/\omega \right)$ and the dimensionless variables $\sigma$ and $\omega_{12}/\omega$.

Our numerical results for the two parameter sets are presented as two phase diagrams in the plane of $\sigma$ and $\omega_{12}/\omega$ in Figs.~\ref{fig:phase_diagram_set1} and~\ref{fig:phase_diagram_set2}. In the remainder of this section, we will first construct the phase diagrams in detail (Sec.~\ref{subsc:phase_diagrams}) and then discuss the emerging phenomena of Rabi suppression~(Sec.~\ref{subsc:phase_suppression}) and domain walls~(Sec.~\ref{subsc:domain_walls}). We recall from Ref.~\cite{PhysRevA.88.013634} that when the Rabi couplings are not present, only triangular lattices were observed in the rotating ground states in the range $0\leq \sigma< 1$. 

\subsection{Lattice phase diagrams}
\label{subsc:phase_diagrams}

\begin{figure}
\includegraphics[width=0.826\columnwidth,keepaspectratio]{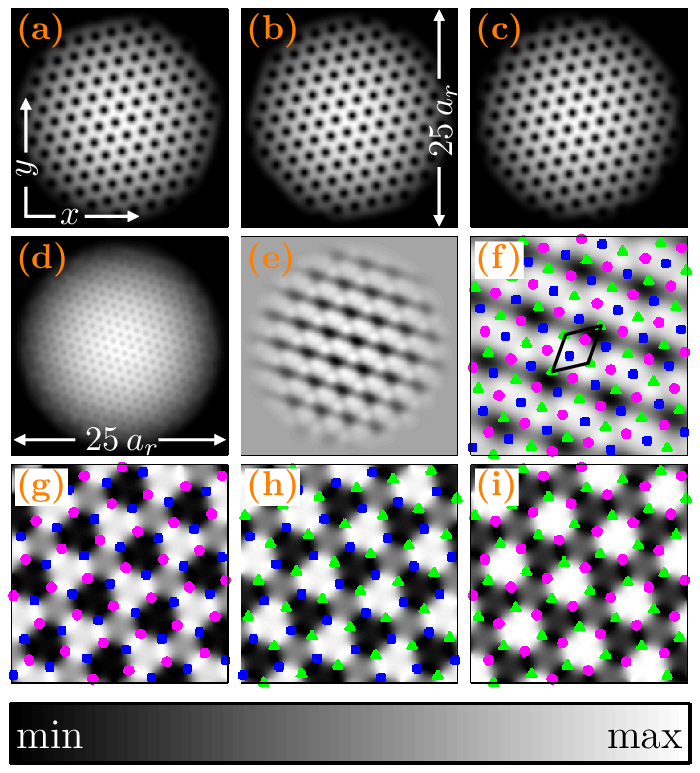}
\caption{\label{fig:s02_w12_001_set1} Rotating ground state for the intercomponent interaction strength $\sigma=g_{12}/g_{11}=0.2$ and the dominant Rabi frequency $\omega_{12}/\omega=-0.01$. The panels show (a)--(c)~atomic densities $|\Psi_1|^2$, $|\Psi_2|^2$, and $|\Psi_3|^2$, respectively; (d)~total density $\ntot$; (e)~negative Rabi energy density $-\rabidensity$; (f)~vortices of each component on top of $-\rabidensity$, with (blue) squares, (magenta) dots, and (green) triangles denoting vortices in the wave functions $\Psi_1$, $\Psi_2$, and $\Psi_3$, respectively; (g)~$-\cos\phase_{12}$, where $\phase_{ij}=\arg\left(\Psi_i\right)-\arg\left(\Psi_j\right)$; (h)~$\cos\phase_{13}$; (i)~$\cos\phase_{23}$. Panel (f) also indicates the elementary unit cell of the combined lattice formed by the three kinds of vortices. The average topological index $\avrtopind=1$ for this state [see Eq.~\eqref{eq:avr_top_ind}]. This state corresponds to the intracomponent interaction strength $g=g_{11} m N/3 \hbar^2=2115$, rotation frequency $\Omega/\omega =0.97$, and Rabi frequencies $\omega_{13}=\omega_{23}=0.01\omega$. The field of view in panels (f)--(i) is $11 a_r\times 11 a_r$, where $a_r=\sqrt{\hbar/m\omega}$, showing the central portion of the harmonic trap. The range of the colormap is from $-1$ to $+1$ in panels (g)--(i), but varies between panels (a)--(f).}
\end{figure}

\begin{figure}
\includegraphics[width=0.826\columnwidth,keepaspectratio]{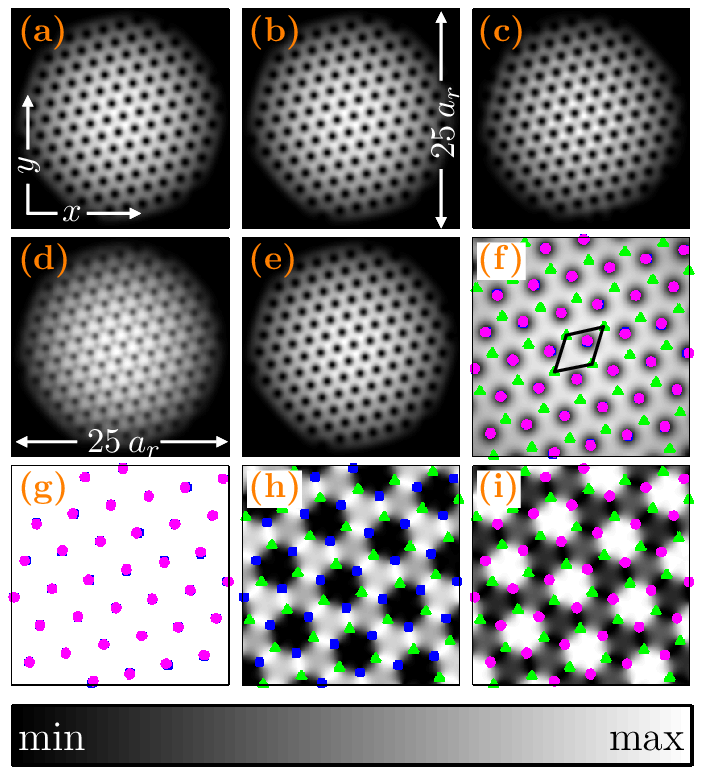}
\caption{\label{fig:s02_w12_009_set1} Rotating ground state for $\sigma=g_{12}/g_{11}=0.2$ and $\omega_{12}/\omega =-0.09$. Other parameter values are the same as in Fig.~\ref{fig:s02_w12_001_set1}. The panels depict (a)~$|\Psi_1| ^2$; (b)~$|\Psi_2| ^2$; (c)~$|\Psi_3| ^2$; (d)~$\ntot$; (e)~$-\rabidensity$; (f)~vortices superposed on $-\rabidensity$; (g)~$-\cos\phase_{12}$; (h)~$\cos\phase_{13}$; (i)~$\cos\phase_{23}$. Panel (f) also indicates the elementary unit cell of the combined vortex lattice. The average topological index $\avrtopind=2/3$ for this state. The field of view in panels (f)--(i) is $11 a_r\times 11 a_r$. The colormap ranges from $0$ to a varying positive maximum in panels (a)--(f) and from $-1$ to $+1$ in panels (g)--(i).}
\end{figure}

\begin{figure}
\includegraphics[width=0.826\columnwidth,keepaspectratio]{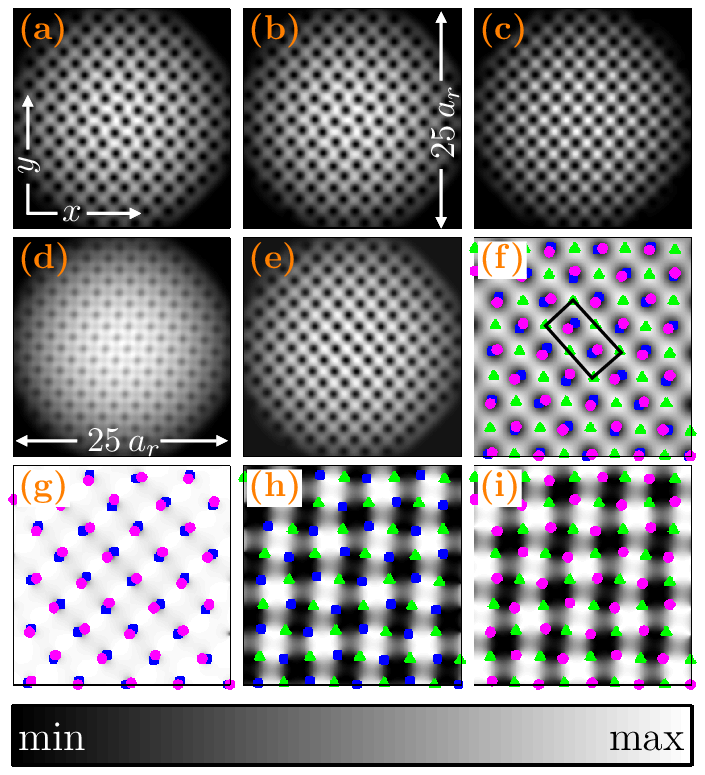}
\caption{\label{fig:s07_w12_009_set1} Rotating ground state for $\sigma=g_{12}/g_{11}=0.7$ and $\omega_{12}/\omega = -0.09$. Other parameter values are the same as in Fig.~\ref{fig:s02_w12_001_set1}. The panels depict (a)~$|\Psi_1| ^2$; (b)~$|\Psi_2| ^2$; (c)~$|\Psi_3| ^2$; (d)~$\ntot$; (e)~$-\rabidensity$; (f)~vortices superposed on $-\rabidensity$; (g)~$-\cos\phase_{12}$; (h)~$\cos\phase_{13}$; (i)~$\cos\phase_{23}$. Panel (f) also shows the elementary unit cell of the combined vortex lattice. The average topological index $\avrtopind=4/3$ for this state. The field of view in panels (f)--(i) is $11 a_r\times 11 a_r$. The colormap ranges from $0$ to a varying positive maximum in panels (a)--(f) and from $-1$ to $+1$ in panels (g)--(i).}
\end{figure}

\begin{figure}
\includegraphics[width=0.826\columnwidth,keepaspectratio]{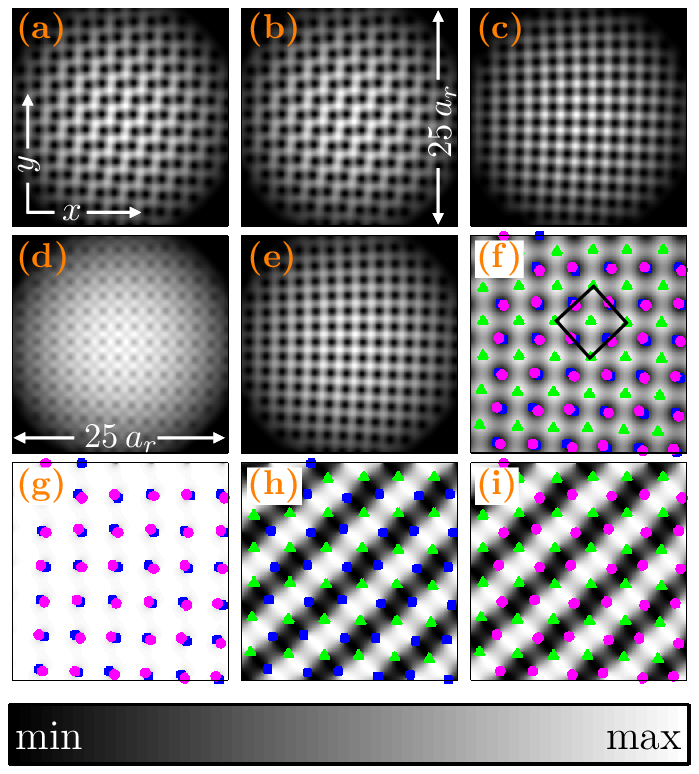}
\caption{\label{fig:s09_w12_009_set1} Rotating ground state for $\sigma=0.9$ and $\omega_{12}/\omega = -0.09$. Other parameter values are the same as in Fig.~\ref{fig:s02_w12_001_set1}. The panels show (a)~$|\Psi_1| ^2$; (b)~$|\Psi_2| ^2$; (c)~$|\Psi_3| ^2$; (d)~$\ntot$; (e)~$-\rabidensity$; (f)~vortices superposed on $-\rabidensity$; (g)~$-\cos\phase_{12}$; (h)~$\cos\phase_{13}$; (i)~$\cos\phase_{23}$. Panel (f) also indicates the elementary unit cell of the combined vortex lattice. The average topological index $\avrtopind=4/3$ for this state. The field of view in panels (f)--(i) is $11 a_r\times 11 a_r$. The colormap ranges from $0$ to a varying positive maximum in panels (a)--(f) and from $-1$ to $+1$ in panels (g)--(i).}
\end{figure}

\begin{figure}
\includegraphics[width=0.826\columnwidth,keepaspectratio]{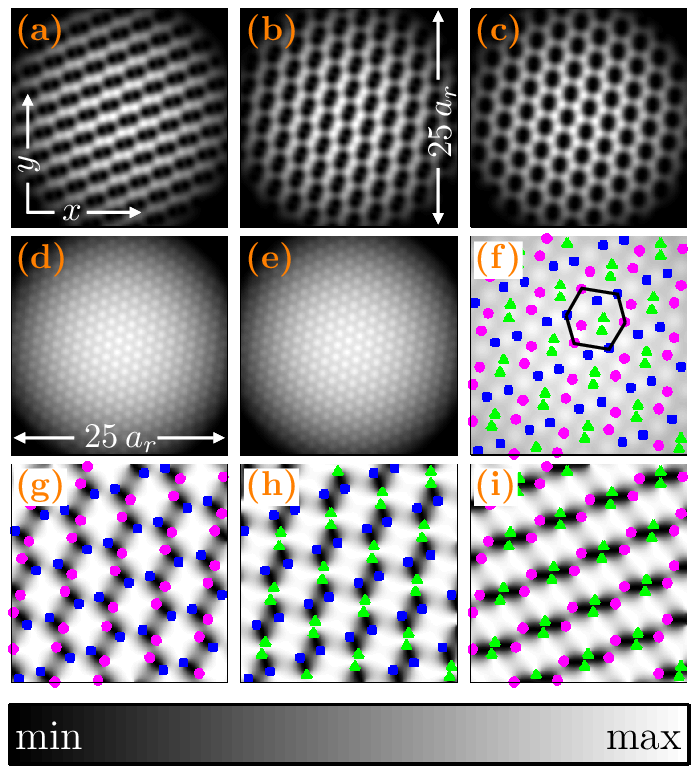}
\caption{\label{fig:s1_w12_001_set1} Rotating ground state for $\sigma=1$ and $\omega_{12}/\omega =-0.01$. Other parameter values are the same as in Fig.~\ref{fig:s02_w12_001_set1}. The panels correspond to (a)~$|\Psi_1| ^2$; (b)~$|\Psi_2| ^2$; (c)~$|\Psi_3| ^2$; (d)~$\ntot$; (e)~$-\rabidensity$; (f)~vortices superposed on $-\rabidensity$; (g)~$-\cos\phase_{12}$; (h)~$\cos\phase_{13}$; (i)~$\cos\phase_{23}$. Panel (f) also shows the elementary unit cell of the combined vortex lattice. The average topological index $\avrtopind=2$ for this state. The field of view in panels (f)--(i) is $11 a_r\times 11 a_r$. The colormap ranges from $0$ to a varying positive maximum in panels (a)--(f) and from $-1$ to $+1$ in panels (g)--(i).}
\end{figure}

\begin{figure}
\includegraphics[width=0.826\columnwidth,keepaspectratio]{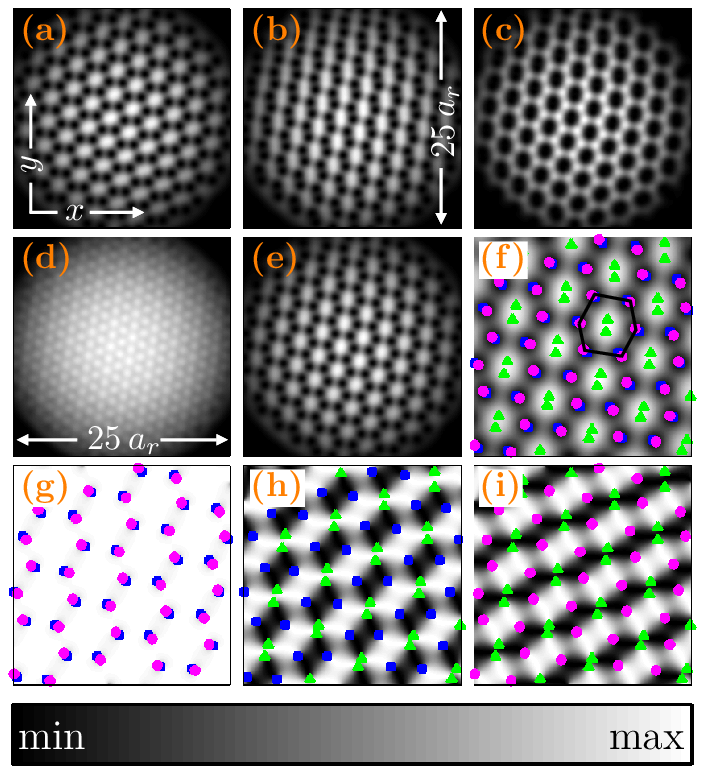}
\caption{\label{fig:s1_w12_009_set1} Rotating ground state for $\sigma=1$ and $\omega_{12}/\omega = -0.09$. Other parameter values are the same as in Fig.~\ref{fig:s02_w12_001_set1}. The panels correspond to (a)~$|\Psi_1| ^2$; (b)~$|\Psi_2| ^2$; (c)~$|\Psi_3| ^2$; (d)~$\ntot$; (e)~$-\rabidensity$; (f)~vortices  superposed on $-\rabidensity$; (g)~$-\cos\phase_{12}$; (h)~$\cos\phase_{13}$; (i)~$\cos\phase_{23}$. Panel (f) also indicates the elementary unit cell of the combined vortex lattice. The average topological index $\avrtopind=4/3$ for this state. The field of view in panels (f)--(i) is $11 a_r\times 11 a_r$. The colormap ranges from $0$ to a varying positive maximum in panels (a)--(f) and from $-1$ to $+1$ in panels (g)--(i).}
\end{figure}

Let us first consider the parameter set with $\omega_{13}=\omega_{23}=0.01\omega$, and vary the interspecies interaction strength and the remaining Rabi frequency in the ranges $0.1\leq \sigma \leq 1$ and $0.01\omega \leq -\omega_{12} \leq 0.12\omega$, respectively. This results in a diverse set of ground-state vortex lattices, examples of which are depicted in Figs.~\ref{fig:s02_w12_001_set1}--\ref{fig:s1_w12_009_set1}. For each solution, we present the density $|\Psi_i|^2$ of each component, the total density $\ntot=\sum_{i}|\Psi_i|^2$, and the Rabi energy density $\rabidensity$; we also present the relative phase angles between the components using the quantities $\mathrm{sgn}\left(\omega_{ij}\right)\cos(\phase_i-\phase_j)$, where $\mathrm{sgn}$ is  the sign function, $i<j$, and $\phase_i=\mathrm{arg}\left(\Psi_i\right)$. In addition, we locate the vortices as the singular points of the superfluid velocity fields $\hbar \nabla \phase_i/m$; the uncertainty in their position is of the order of the grid spacing, which is $0.0875 a_r$ throughout this work. We will refer to Figs.~\ref{fig:s02_w12_001_set1}--\ref{fig:s1_w12_009_set1} when discussing the related phenomena in the subsequent sections. 

For $0.1\leq \sigma \leq 0.9$ and sufficiently small values of $|\omega_{12}|$, all three components host triangular vortex lattices that are interlaced with one another. An example of such a state is presented in Fig.~\ref{fig:s02_w12_001_set1} for $\sigma=0.2$ and $\omega_{12}=-0.01\omega$. However, for $0.1 \leq \sigma \leq 0.2$ and increased $|\omega_{12}|$, vortices in components~1 and~2 move on top of each other to form overlapping triangular vortex lattices, which are in turn interlaced by the triangular lattice in component~3~(Fig.~\ref{fig:s02_w12_009_set1}). Together, the three components constitute a honeycomb lattice of local minima in the total density $\ntot$ [Fig.~\ref{fig:s02_w12_009_set1}(d)].

For stronger intercomponent repulsion within the miscible regime, $0.3 \leq \sigma \leq 0.9$, and increased $|\omega_{12}|$, the triangular vortex lattices in components~1 and~2 become replaced by almost overlapping square lattices of \emph{vortex dimers}, while component~3 hosts a square lattice of solitary vortices that interlaces both dimer lattices. Ground states of this type are shown in Figs.~\ref{fig:s07_w12_009_set1} and~\ref{fig:s09_w12_009_set1}. In the range $0.3 \leq \sigma \leq 0.7$, the alignment of the dimers tends to exhibit small distortions across the system, as is evident from Figs.~\ref{fig:s07_w12_009_set1}(a) and~\ref{fig:s07_w12_009_set1}(b). Note, however, that neither $\ntot$ [Fig.~\ref{fig:s07_w12_009_set1}(d)] nor the Rabi energy density $\rabidensity$~[Fig.~\ref{fig:s07_w12_009_set1}(e)] shows the lattice distortions appearing in components~1 and~2. For $0.8 \leq \sigma \leq 0.9$, on the other hand, the dimers tend to be uniformly aligned, as in Figs.~\ref{fig:s09_w12_009_set1}(a) and~\ref{fig:s09_w12_009_set1}(b).

The case $\sigma=1$, corresponding to strong intercomponent repulsion, can be considered as the border that separates miscible and immiscible regimes. Here, two lattice phases can be distinguished with varying $\omega_{12}$. For $0.01\omega \leq -\omega_{12} \leq 0.04\omega$, we obtain interlaced triangular vortex-dimer lattices in components~1 and~2, while a triangular lattice of doubly quantized fused-core vortices appears in component~3 (Fig.~\ref{fig:s1_w12_001_set1}). A fused-core vortex comprises two singly quantized vortices practically coalesced into one doubly quantized defect, or at least to within a distance  smaller than the core diameter of the constituent vortices. The appearance of doubly quantized vortices in the ground state of the system exemplifies the versatile rotational behavior of multicomponent BECs and is to be contrasted with single-component condensates, where multiply quantized vortices tend to be highly unstable against splitting~\cite{Pu1999.PRA59.1533,Kaw2004.PRA70.043610,Kuo2010.PRA81.023603,Kuo2010.PRA81.033627} unless specifically stabilized by confinement~\cite{Kuo2010.PRA81.033627,Kuo2010.JLTP161.561,Sim2002.PRA65.033614}. The distinct elliptical shape of the combined defect [Fig.~\ref{fig:s1_w12_001_set1}(c)]  is due to the small separation of the phase singularities within the fused core. In the second lattice phase at $\sigma=1$, which occurs for $0.05\omega \leq -\omega_{12} \leq 0.12\omega$ and is illustrated in Fig.~\ref{fig:s1_w12_009_set1}, component~1 hosts a honeycomb vortex lattice, component~2 a triangular lattice of vortex dimers, and component~3 a triangular lattice of fused-core vortices. Furthermore, we note that the Rabi energy density exhibits a honeycomb spatial structure [Fig.~\ref{fig:s1_w12_009_set1}(e)]. 

\begin{figure}
\includegraphics[width=0.987\columnwidth,keepaspectratio]{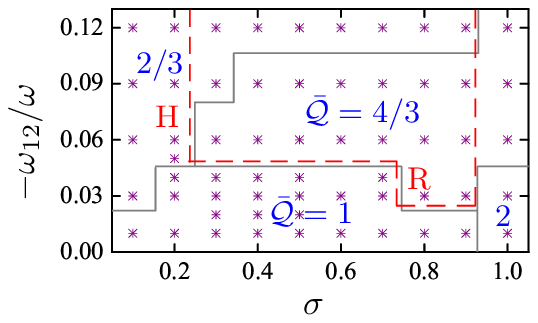}
\caption{\label{fig:phase_diagram_set1}Phase diagram of skyrmionic vortex lattices as a function of the intercomponent interaction strength $\sigma$ and the Rabi frequency $\omega_{12}$, for fixed intracomponent interaction strength $g=2115$, external rotation frequency $\Omega/\omega =0.97$,  and Rabi frequencies $\omega_{13}=\omega_{23}=0.01\omega$. Each asterisk corresponds to a numerically solved ground state of the rotating three-component Bose--Einstein condensate. The gray solid lines demarcate regions with different indicated values of the average topological index $\avrtopind$ [Eq.~\eqref{eq:avr_top_ind}]. The (red) dashed lines indicate the boundary across which the geometry of the elementary unit cell of the three-component vortex lattice changes from hexagonal (H) to rectangular (R) or vice versa.}
\end{figure}

All the states discussed above can actually be topologically characterized as containing skyrmions, which have attracted considerable attention in the context of multicomponent BECs~\cite{Alkhawaja.nature411,PhysRevA.77.033621,PhysRevA.84.033611,PhysRevA.91.043605,Eto2013.EPL103.60006,Wu2011.ChinPhysLett28.097102,Zho2011.PRA84.063624,Li2016.PRA93.033628,Cho2012.PRL108.035301,Oll2014.PRA89.033629}.  As detailed in Ref.~\cite{PhysRevB.87.014507}, skyrmions in a $K$-component model in two spatial dimensions can be defined by the $\mathbb{C} P^{K-1}$ topological invariant 
\begin{equation}
\topind=\int \frac{i \epsilon_{\beta\alpha}}{2\pi |\Psi|^4} \Big( |\Psi|^2  \partial_\alpha \Psi^\dagger \partial_\beta \Psi + \Psi^\dagger \partial_\alpha \Psi \partial_\beta \Psi^\dagger \Psi \Big) d^2 r,
\label{eq:top_ind}
\end{equation}
termed the topological index. $\mathbb{C}P^{K-1}$ is the complex projective space whose points label the complex lines through the origin of the space $\mathbb{C}^{K}$. In Eq.~\eqref{eq:top_ind}, $\Psi^\dagger=(\Psi^\ast_1,\Psi^\ast_2,\dots,\Psi^\ast_K)$, $\epsilon_{\beta\alpha}$ is the two-dimensional Levi-Civita symbol, and summation over $\alpha,\beta\in\left\{x,y\right\}$ is implied. For our states, the integration in Eq.~\eqref{eq:top_ind} is carried over an elementary unit cell of the combined three-component vortex lattice, which we determine from the central region of the trap by treating the vortex array as a system of three types of point particles. The chosen elementary unit cells are indicated by the black solid lines in panels~(f) of Figs.~\ref{fig:s02_w12_001_set1}--\ref{fig:s1_w12_009_set1}. The unit cell is also used to categorize the overall lattice geometry as either rectangular or hexagonal: we identify the geometry as rectangular if the elementary unit cell can be chosen so that its largest angle is closer to $\pi/2$ than it is to $2\pi/3$; otherwise, the geometry is identified as hexagonal~\cite{hexagonal_note}. We note in passing that, from a topological point of view, the existence of two-dimensional skyrmions is allowed  in the model because the second homotopy group $\pi_2\left(\mathbb{C}P^{K-1}\right)$ for $K\geq 2$ is isomorphic to the additive group of integers, $\mathbb{Z}$~\cite{Mer1979.RMP51.591,Hin1993.NuclPhysB392.461}.

The topological index $\topind$ is zero for an integer vortex, i.e., when there is an equally charged vortex in every component at the same point in space (or, in fact, when the vortices are separated by distances significantly smaller than their core radii). In our case, we deal with two types of states: In the first type, which corresponds to small values of $|\omega_{12}|$, there are three mutually interlaced lattices, as, for example, in Fig.~\ref{fig:s02_w12_001_set1}. In the second type, which appears when $|\omega_{12}|$ is increased, we find increasingly overlapping vortex lattices in components~1 and~2, as in Fig.~\ref{fig:s02_w12_009_set1}. If we calculate the $\mathbb{C} P^2$ topological index for the states in Figs.~\ref{fig:s02_w12_001_set1} and~\ref{fig:s02_w12_009_set1}, we obtain the same $\topind=1$ per combined-lattice unit cell, although these two examples clearly constitute two distinct phases. Therefore, in order to better distinguish between different phases, we instead calculate pairwise $\mathbb{C} P^1$ topological indices $\topind_{ij}$, $i<j$, by using Eq.~\eqref{eq:top_ind} separately for each pair of the components~\cite{Qij_note}, and then calculate an average topological index $\avrtopind$ for the entire three-component state as
\begin{equation}
\avrtopind=\frac{\topind_{12}+\topind_{13}+\topind_{23}}{3}.
\label{eq:avr_top_ind}
\end{equation}
We stress that each $\topind_{ij}$ is calculated over the same unit cell of the \emph{three-component} lattice. In all the cases we consider here, $\topind_{ij}\in\left\{0,1,2\right\}$.

To illustrate the use of Eq.~\eqref{eq:avr_top_ind}, let us consider, for example, the state shown in Fig.~\ref{fig:s02_w12_001_set1}, for which the elementary unit cell is chosen as the rhombus that connects four vortices of component~3 [Fig.~\ref{fig:s02_w12_001_set1}(f)]. The unit cell encloses one vortex of each component. As a result, the pairwise topological indices are $\topind_{12}=\topind_{13}=\topind_{23}=1$, and hence the average topological index for the state is $\avrtopind=1$. On the other hand, calculating the average topological index for the state  shown in Fig.~\ref{fig:s02_w12_009_set1} in the same manner gives $\avrtopind=2/3$, because the individual vortices of components~1 and~2 reside on top of each other and hence $\topind_{12}=0$.  

By calculating the average topological index $\avrtopind$ from Eq.~\eqref{eq:avr_top_ind} for all the obtained states and collecting the results, we obtain the phase diagram shown in Fig.~\ref{fig:phase_diagram_set1}. It classifies the different types of ground-state skyrmionic vortex lattices for the fixed Rabi couplings $\omega_{13}=\omega_{23}=0.01\omega$ in the two-dimensional domain $0.1 \leq \sigma \leq 1$ and $0.01\omega \leq -\omega_{12} \leq 0.12\omega$. In addition to indicating the changes in $\avrtopind$, Fig.~\ref{fig:phase_diagram_set1} shows the boundary across which the geometry of the combined vortex lattice changes from hexagonal to rectangular or vice versa. 

The states with $\avrtopind=1$ correspond to skyrmionic lattices with one vortex per component in a rhombic unit cell that results from three mutually interlaced triangular vortex lattices [Fig.~\ref{fig:s02_w12_001_set1}(f)]. This phase occurs for $\sigma \leq 0.9$ and small $|\omega_{12}|$. For $\sigma \leq 0.2$ and large $|\omega_{12}|$, the states with $\avrtopind = 2/3$ also have a rhombic unit cell but with overlapping triangular lattices in components~1 and~2 [Fig.~\ref{fig:s02_w12_009_set1}(f)]. However, for $\sigma \geq 0.3$, the states with $\avrtopind=2/3$ are described by a square unit cell instead. The rectangular lattice phase with $\avrtopind = 4/3$, which occurs for $0.3 \leq \sigma \leq 0.9$ at intermediate values of $|\omega_{12}|$, originates from almost overlapping square lattices of vortex dimers in components~1 and~2 that are interlaced by a square vortex lattice in component~3, implying a rectangular unit cell that contains two vortices of each component [Figs.~\ref{fig:s07_w12_009_set1}(f) and~\ref{fig:s09_w12_009_set1}(f)]. The states with $\avrtopind=2$ appearing at $\sigma=1$ for small values of $|\omega_{12}|$ have a hexagonal unit cell that includes two vortices of each component [Fig.~\ref{fig:s1_w12_001_set1}(f)]; with increasing $|\omega_{12}|$, the vortices of component~1 begin to coincide with those of component~2, resulting in a hexagonal lattice phase with $\avrtopind=4/3$ for $\sigma =1$ and $-\omega_{12} \geq 0.06 \omega$ [Fig.~\ref{fig:s1_w12_009_set1}(f)].

\begin{figure}
\includegraphics[width=0.826\columnwidth,keepaspectratio]{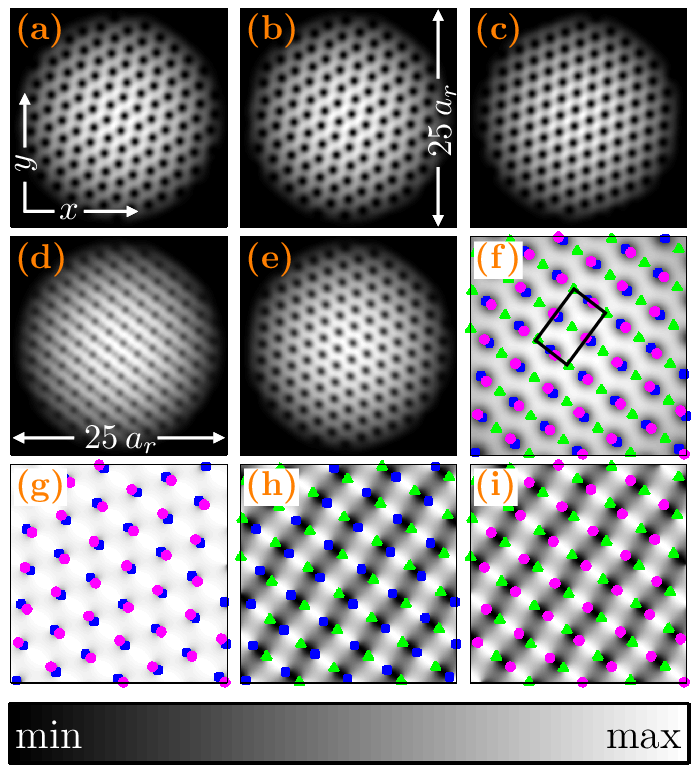}
\caption{\label{fig:s02_w12_009_set2} Rotating ground state for the intercomponent interaction strength $\sigma=g_{12}/g_{11}=0.2$ and the dominant Rabi frequency $\omega_{12}/\omega=-0.09$. The panels show (a)--(c)~atomic densities $|\Psi_1|^2$, $|\Psi_2|^2$, and $|\Psi_3|^2$, respectively; (d)~total density $\ntot$; (e)~negative Rabi energy density $-\rabidensity$; (f)~vortices of each component on top of $-\rabidensity$; (g)~$-\cos\phase_{12}$; (h)~$\cos\phase_{13}$; (i)~$\cos\phase_{23}$. Panel (f) also shows the elementary unit cell of the combined vortex lattice. The average topological index $\avrtopind=4/3$ for this state. This state corresponds to the intracomponent interaction strength $g=g_{11} m N/3 \hbar^2=2115$, rotation frequency $\Omega/\omega =0.97$, and Rabi frequencies $\omega_{13}/\omega=\omega_{23}=0.05\omega$. The field of view in panels (f)--(i) is $11 a_r\times 11 a_r$. The colormap ranges from $0$ to a varying positive maximum in panels (a)--(f) and from $-1$ to $+1$ in panels (g)--(i).}
\end{figure}

\begin{figure}
\includegraphics[width=0.826\columnwidth,keepaspectratio]{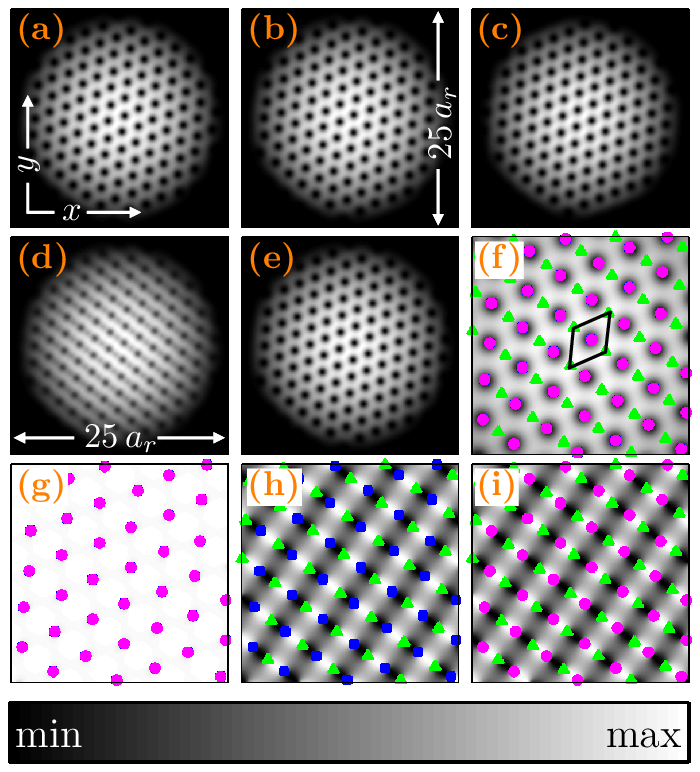}
\caption{\label{fig:s01_w12_016_set2} Rotating ground state for $\sigma=0.1$ and $\omega_{12}/\omega = -0.16$. Other parameter values are the same as in Fig.~\ref{fig:s02_w12_009_set2}. The panels show (a)~$|\Psi_1| ^2$; (b)~$|\Psi_2| ^2$; (c)~$|\Psi_3| ^2$; (d)~$\ntot$; (e)~$-\rabidensity$; (f)~vortices superposed on $-\rabidensity$; (g)~$-\cos\phase_{12}$; (h)~$\cos\phase_{13}$; (i)~$\cos\phase_{23}$. Panel (f) also indicates the elementary unit cell of the combined vortex lattice. The average topological index $\avrtopind=2/3$ for this state. The field of view in panels (f)--(i) is $11 a_r\times 11 a_r$. The colormap ranges from $0$ to a varying positive maximum in panels (a)--(f) and from $-1$ to $+1$ in panels (g)--(i).}
\end{figure}

\begin{figure}
\includegraphics[width=0.826\columnwidth,keepaspectratio]{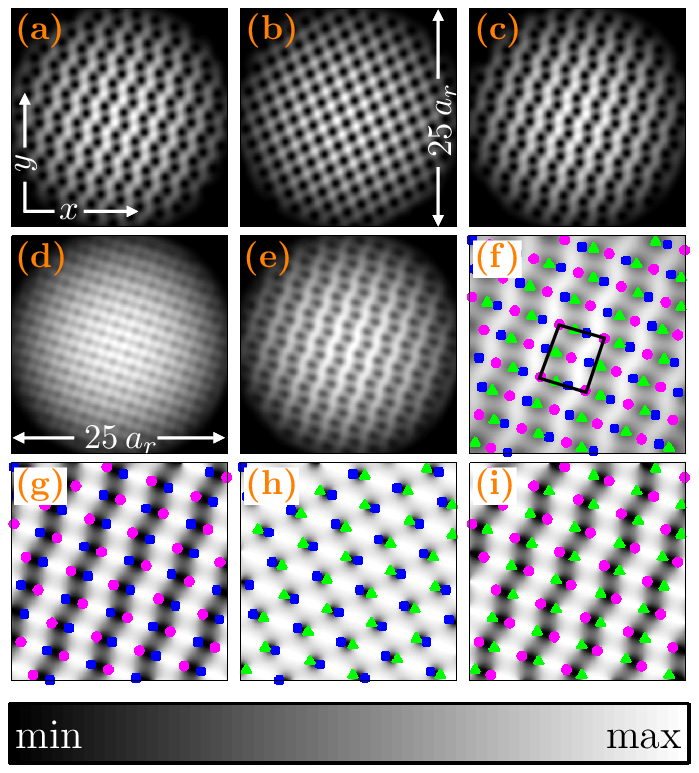}
\caption{\label{fig:s07_w12_003_set2} Rotating ground state for $\sigma=0.7$ and $\omega_{12}/\omega = -0.03$. Other parameter values are the same as in Fig.~\ref{fig:s02_w12_009_set2}. The panels show (a)~$|\Psi_1| ^2$; (b)~$|\Psi_2| ^2$; (c)~$|\Psi_3| ^2$; (d)~$\ntot$; (e)~$-\rabidensity$; (f)~vortices superposed on $-\rabidensity$; (g)~$-\cos\phase_{12}$; (h)~$\cos\phase_{13}$; (i)~$\cos\phase_{23}$. Panel (f) also indicates the elementary unit cell of the combined vortex lattice. The average topological index $\avrtopind=2$ for this state. The field of view in panels (f)--(i) is $11 a_r\times 11 a_r$. The colormap ranges from $0$ to a varying positive maximum in panels (a)--(f) and from $-1$ to $+1$ in panels (g)--(i).}
\end{figure}

\begin{figure}
\includegraphics[width=0.826\columnwidth,keepaspectratio]{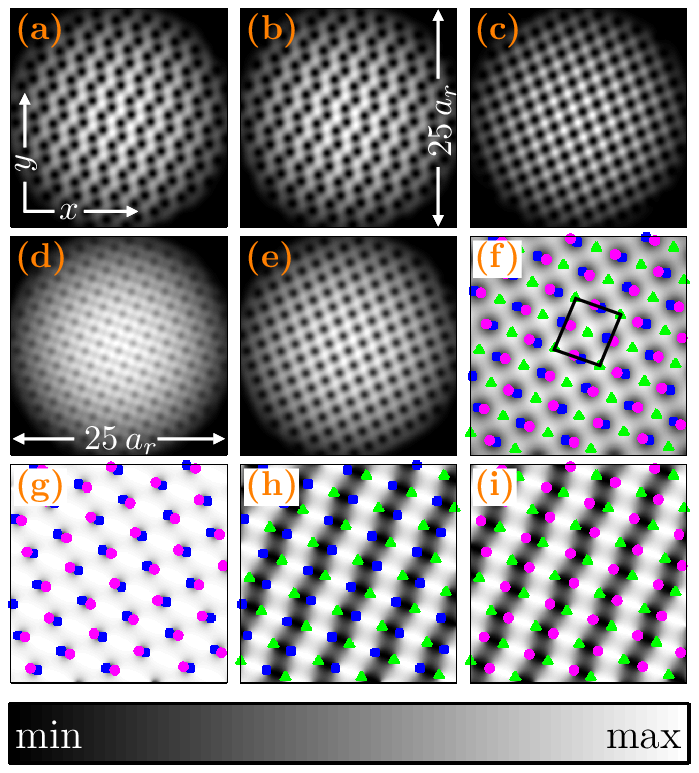}
\caption{\label{fig:s07_w12_012_set2} Rotating ground state for $\sigma=0.7$ and $\omega_{12}/\omega =-0.12$. Other parameter values are the same as in Fig.~\ref{fig:s02_w12_009_set2}. The panels show (a)~$|\Psi_1| ^2$; (b)~$|\Psi_2| ^2$; (c)~$|\Psi_3| ^2$; (d)~$\ntot$; (e)~$-\rabidensity$; (f)~vortices superposed on $-\rabidensity$; (g)~$-\cos\phase_{12}$; (h)~$\cos\phase_{13}$; (i)~$\cos\phase_{23}$. Panel (f) also shows the elementary unit cell of the combined vortex lattice. The average topological index $\avrtopind=4/3$ for this state. The field of view in panels (f)--(i) is $11 a_r\times 11 a_r$. The colormap ranges from $0$ to a varying positive maximum in panels (a)--(f) and from $-1$ to $+1$ in panels (g)--(i).}
\end{figure}

\begin{figure}
\includegraphics[width=0.826\columnwidth,keepaspectratio]{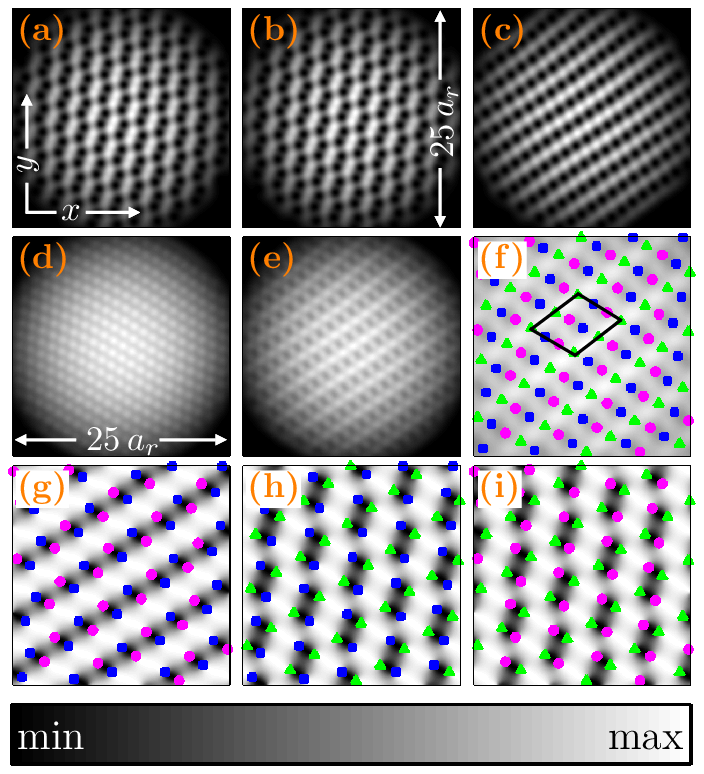}
\caption{\label{fig:s09_w12_006_set2} Rotating ground state for $\sigma=0.9$ and $\omega_{12}/\omega = -0.06$. Other parameter values are the same as in Fig.~\ref{fig:s02_w12_009_set2}. The panels depict (a)~$|\Psi_1| ^2$; (b)~$|\Psi_2| ^2$; (c)~$|\Psi_3| ^2$; (d)~$\ntot$; (e)~$-\rabidensity$; (f)~vortices superposed on $-\rabidensity$; (g)~$-\cos\phase_{12}$; (h)~$\cos\phase_{13}$; (i)~$\cos\phase_{23}$. Panel (f) also shows the elementary unit cell of the combined vortex lattice. The average topological index $\avrtopind=2$ for this state.  The field of view in panels (f)--(i) is $11 a_r\times 11 a_r$. The colormap ranges from $0$ to a varying positive maximum in panels (a)--(f) and from $-1$ to $+1$ in panels (g)--(i).}
\end{figure}

\begin{figure}
\includegraphics[width=0.826\columnwidth,keepaspectratio]{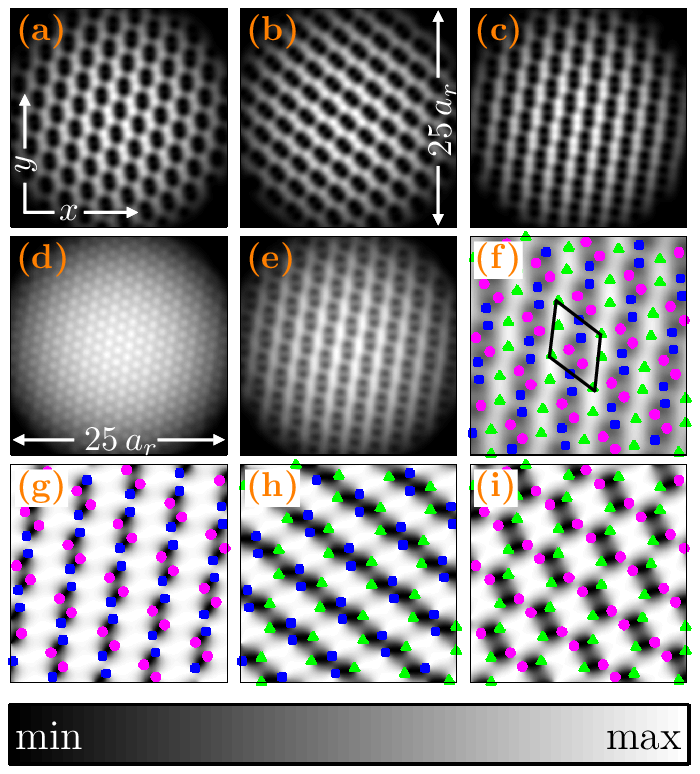}
\caption{\label{fig:s1_w12_009_set2} Rotating ground state for $\sigma=1$ and $\omega_{12}/\omega = -0.09$. Other parameter values are the same as in Fig.~\ref{fig:s02_w12_009_set2}. The panels show (a)~$|\Psi_1| ^2$; (b)~$|\Psi_2| ^2$; (c)~$|\Psi_3| ^2$; (d)~$\ntot$; (e)~$-\rabidensity$; (f)~vortices superposed on $-\rabidensity$; (g)~$-\cos\phase_{12}$; (h)~$\cos\phase_{13}$; (i)~$\cos\phase_{23}$. Panel (f) also shows the elementary unit cell of the combined vortex lattice. The average topological index $\avrtopind=2$ for this state. The field of view in panels (f)--(i) is $11 a_r\times 11 a_r$. The colormap ranges from $0$ to a varying positive maximum in panels (a)--(f) and from $-1$ to $+1$ in panels (g)--(i).}
\end{figure}

\begin{figure}
\includegraphics[width=0.974\columnwidth,keepaspectratio]{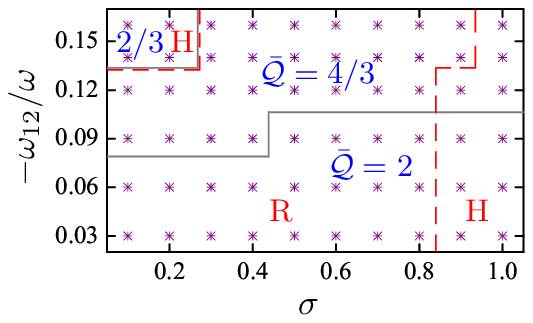}
\caption{\label{fig:phase_diagram_set2} Phase diagram of skyrmionic vortex lattices in the plane of the intercomponent interaction strength $\sigma$ and the Rabi frequency $\omega_{12}$ for fixed intracomponent interaction strength $g=2115$, external rotation frequency $\Omega/\omega =0.97$, and Rabi frequencies $\omega_{13}=\omega_{23}=0.05\omega$. Each asterisk corresponds to a numerically solved ground state of the three-component Bose--Einstein condensate. The gray solid lines demarcate regions with different indicated values of the average topological index $\avrtopind$ [Eq.~\eqref{eq:avr_top_ind}]. The (red) dashed lines mark the boundary across which the geometry of the elementary unit cell of the three-component vortex lattice changes from to rectangular (R) to hexagonal (H) or vice versa.}
\end{figure}

We now turn to the second parameter set, which differs from the first by having $\omega_{13}=\omega_{23}=0.05\omega$ instead of $0.01\omega$. Representative ground-state solutions for $0.03\omega \leq -\omega_{12} \leq 0.16\omega$ and $0.1 \leq \sigma \leq 1$ are depicted in Figs.~\ref{fig:s02_w12_009_set2}--\ref{fig:s1_w12_009_set2}. When all the obtained solutions from this range are classified in terms of the average topological index $\avrtopind$ [Eq.~\eqref{eq:avr_top_ind}], we obtain the skyrmionic-lattice phase diagram presented in Fig.~\ref{fig:phase_diagram_set2}. In addition to indicating the observed values of $\avrtopind$, the diagram categorizes the ground states as hexagonal or rectangular according to the geometry of the elementary unit cell. Below, we provide a detailed account of the discovered phases.

We first consider the small-$\sigma$ regime (Figs.~\ref{fig:s02_w12_009_set2} and~\ref{fig:s01_w12_016_set2}). For small $|\omega_{12}|$,  components~1 and~2 host zig-zag vortex lattices and component~3 exhibits a conventional triangular lattice. All three lattices interlace one another, and the unit cell of the combined vortex lattice is a rectangle containing two vortices of each component; hence, the average topological index is $\avrtopind=2$. With increasing $|\omega_{12}|$,  the zig-zag lattices in components~1 and~2 begin to overlap more and more, while the triangular lattice in component~3 remains interlaced with the other two, giving rise to $\avrtopind=4/3$ (Fig.~\ref{fig:s02_w12_009_set2}). In the limit of large $|\omega_{12}|$, the lattices in components~1 and~2 become locked together and lose their zig-zag character, which results in a two-component-like phase with $\avrtopind=2/3$ and a rhombic unit cell~(Fig.~\ref{fig:s01_w12_016_set2}). The total density exhibits a plane-wave-like modulation in both Figs.~\ref{fig:s02_w12_009_set2}(d) and~\ref{fig:s01_w12_016_set2}(d).  

The states occurring for intermediate strengths of the intercomponent repulsion, $0.5 \lesssim \sigma \lesssim 0.8$, correspond to two components hosting zig-zag lattices and one component having a square lattice. For small $|\omega_{12}|$, all three lattices interlace one another and $\avrtopind=2$ (Fig.~\ref{fig:s07_w12_003_set2}). With increasing $|\omega_{12}|$, the separation distance between component-1 and component-2 vortices decreases and the average topological index eventually falls to $\avrtopind=4/3$ (Fig.~\ref{fig:s07_w12_012_set2}). In both phases, the unit cell is a rectangle that encloses two vortices of each component and approaches a square with increasing $|\omega_{12}|$ [Figs.~\ref{fig:s07_w12_003_set2}(f) and~\ref{fig:s07_w12_012_set2}(f)]. We also note that while the spatial profile of the Rabi energy density $\rabidensity$ depends noticeably on $\omega_{12}$  [Figs.~\ref{fig:s07_w12_003_set2}(e) and~\ref{fig:s07_w12_012_set2}(e)], the total density $\ntot$ exhibits a square pattern in both phases  [Figs.~\ref{fig:s07_w12_003_set2}(d) and~\ref{fig:s07_w12_012_set2}(d)].

When the intercomponent interaction strength $\sigma$ approaches unity, the overall geometry of the three-component lattices changes back to hexagonal. For $\sigma=0.9$ and $0.03\omega \leq -\omega_{12}\leq 0.09\omega$, the system hosts triangular dimer lattices in components~1 and~2 and parallel straight chains of vortices in component~3, as illustrated in Fig.~\ref{fig:s09_w12_006_set2} for $\omega_{12}=-0.06\omega$. For $\sigma=1$ and $0.03\omega \leq -\omega_{12}\leq 0.09\omega$, the dimers in components~1 and~2 become more tightly bound and turn into fused-core vortices, while component~3 hosts a triangular lattice of dimers (Fig.~\ref{fig:s1_w12_009_set2}). As shown in Figs.~\ref{fig:s09_w12_006_set2}(f) and~\ref{fig:s1_w12_009_set2}(f), the unit cell for both of these states is a rhomboid containing two vortices of each component, and the average topological index is $\avrtopind=2$. With increasing $|\omega_{12}|$, the vortices in component~1 move on top of those in component~2, resulting in the hexagonal lattice phase with $\avrtopind=4/3$ observed for $0.12 \omega \leq -\omega_{12} \leq 0.16\omega$.  

\subsection{Rabi suppression in the three-component system}
\label{subsc:phase_suppression}

\begin{figure}
\includegraphics[width=0.919822\columnwidth,keepaspectratio]{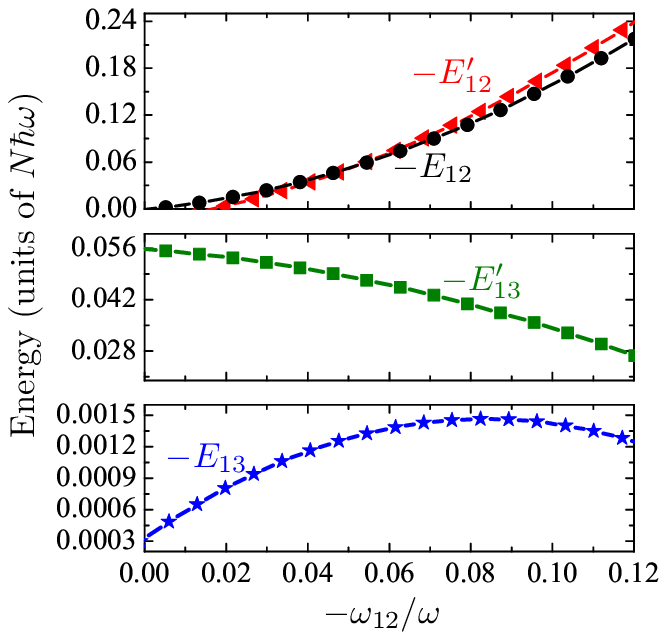}
\caption{\label{fig:rabi} Pairwise Rabi energies $E_{ij}=-\hbar \omega_{ij}\int ( \Psi_i^{\ast}\Psi_j + \Psi_j^{\ast} \Psi_i )\,d^2 r$ as functions of the Rabi frequency $\omega_{12}$ for fixed $\sigma=0.7$, $g=2115$, and $\Omega/\omega =0.97$. Here $E_{ij}$ are for $\omega_{13}=\omega_{23}=0.01\omega$ and $E^{\prime}_{ij}$ for $\omega_{13}=\omega_{23}=0.05\omega$.}
\end{figure}

In the Gross--Pitaevskii model for rotating Rabi-coupled \emph{two-component} BECs, which is obtained from Eq.~\eqref{eq:func_en} by assuming $\Psi_3\equiv 0$, the sign of the Rabi frequency $\omega_{12}$ is irrelevant for the ground-state energetics because changing the sign of $\omega_{12}$ can be exactly balanced by changing the sign of either  $\Psi_1$ or $\Psi_2$. Therefore, $\omega_{12}= \omega_0$ and $\omega_{12}=-\omega_0$ ($\omega_0\in\mathbb{R}$) will yield physically identical ground-state solutions with the same attractive Rabi energy $E_\mathrm{R} \leq 0$ that favors coincidence of same-sign vortices between the two components. In the three-component counterpart, however, the signs of $\omega_{ij}$ make a difference, and can result in intrinsic frustration and consequent suppression of some or all of the three pairwise Rabi couplings.

In order to heuristically see how the Rabi suppression emerges in the three-component BEC, consider the wave functions in the vicinity of a vortex, for example, in component~1. In local polar coordinates $\left(r',\phi'\right)$ with the vortex at $r'=0$, we write the wave functions as $\Psi_j\left(r',\phi'\right)=\exp\left[i\left(\kappa_j\phi'+C_j\right)\right]f_k\left(r'\right)$,
where the constants $C_j\in\mathbb{R}$ only affect the Rabi term in Eq.~\eqref{eq:func_en}. The Rabi energy density then becomes
\begin{equation}
\frac{E^\mathrm{loc}_\mathrm{R}}{\pi r_0^2}  = -\frac{2}{r_0^2} \sumsum_{i<j} \omega_{ij} \delta_{\kappa_i, \kappa_j} \cos C_{ij} \int_0^{r_0}  f_if_j  r' dr',
\label{eq:rabi}
\end{equation}
where $\delta_{\kappa_i, \kappa_j}$ is the Kronecker delta, $C_{ij}=C_i-C_j$, and $r_0$ defines the small disk over which the local Rabi energy $E_\mathrm{R}^\mathrm{loc}$ is averaged. In the case $\kappa_1=\kappa_2=\kappa_3=\kappa$, i.e., a $\kappa$-quantum integer vortex (or no vortices at all if $\kappa=0$), all three terms in the sum can be nonzero. If we further assume $f_1=f_2=f_3$, the minimization of the above Rabi energy density implies maximization of the function $h\left(C_{12},C_{13}\right)=\sumsum_{i<j}\omega_{ij}\cos C_{ij}=\omega_{12}\cos C_{12} + \omega_{13}\cos C_{13} + \omega_{23} \cos\left(C_{12}-C_{13}\right)$ with respect to $C_{12}$ and $C_{13}$. This function has an upper bound of $\sumsum_{i<j}|\omega_{ij}|$. However, depending on the values of $\omega_{ij}$, $\max_{C_{ij}}h\left(C_{12},C_{13}\right)$ may be significantly below this upper bound, indicating that some or all of the Rabi couplings are suppressed by the relative phase frustration between the particular components. In general, the upper bound can be reached if and only if $\omega_{12}\omega_{13}\omega_{23} \geq 0$, which is never satisfied by the parameter values used in this work (all the presented states have $\omega_{12} < 0 < \omega_{13}=\omega_{23}$).

Of course, the above calculation based on Eq.~\eqref{eq:rabi}  is only a crude approximation to the intricate behavior we have obtained from the full GP equations~\eqref{eq:GP_cast}. But at least it shows that even with fully overlapping vortex lattices in all three components, corresponding to nonskyrmionic states with $\avrtopind=0$, the amount of energy gained by minimizing the Rabi energy would be strongly suppressed by the inherent phase frustration; this in turn explains why the interlacing of vortices and the consequently rich skyrmionic-lattice and relative-phase structures feature so prominently in the system. The approximation suggests that for the parameter set used in Fig. \ref{fig:phase_diagram_set1}, the phase frustration would occur symmetrically between all three pairs at $\omega_{12}=-0.01\omega$, for which the above maximization yields $\cos C_{12}=-0.5$ and $\cos C_{13}=\cos C_{23}=0.5$ (Figs.~\ref{fig:s02_w12_001_set1} and~\ref{fig:s1_w12_001_set1} show examples with such $\omega_{12}$). At $\omega_{12}=-0.09\omega$~(Figs.~\ref{fig:s02_w12_009_set1}--\ref{fig:s09_w12_009_set1} and~\ref{fig:s1_w12_009_set1}), the optimal relative phases would yield $\cos C_{12}=-0.99$ and $\cos C_{13}=\cos C_{23}=0.056$, so that $\max_{C_{ij}}h\left(C_{12},C_{13}\right)= 0.091 \omega < 0.11 \omega = \sumsum_{i<j}|\omega_{ij}|$. This means that the Rabi couplings within the pairs~1--3 and~2--3 would be strongly suppressed, whereas the coupling within the pair~1--2 would be almost maximal. The prediction is in line with Figs.~\ref{fig:s02_w12_009_set1}--\ref{fig:s09_w12_009_set1} and~\ref{fig:s1_w12_009_set1}, where locking of the vortex lattices is observed within the pair~1--2 but not within the other two pairs. For the parameter set used in Fig.~\ref{fig:phase_diagram_set2}, the phase frustration is expected to be symmetric between all three pairs at $\omega_{12}=-0.05\omega$ (cf. Fig.~\ref{fig:s09_w12_006_set2}, where $\omega_{12}=-0.06\omega$) and to occur dominantly within the pairs~1--3 and~2--3 for $\omega_{12} \leq -0.09\omega$ (Figs.~\ref{fig:s02_w12_009_set2}, \ref{fig:s01_w12_016_set2}, \ref{fig:s07_w12_012_set2}, and~\ref{fig:s1_w12_009_set2}).

Figure~\ref{fig:rabi} shows the behavior of the pairwise Rabi energies  $E_{12}$ and $E_{13}$ as functions of $-\omega_{12}$  for fixed $\sigma=0.7$, $g=2115$, and $\Omega/\omega =0.97$, as obtained from the numerical solution of Eqs.~\eqref{eq:GP_cast}. Unprimed quantities are for $\omega_{13}=\omega_{23}=0.01\omega$ and primed quantities for  $\omega_{13}=\omega_{23}=0.05\omega$. We observe that  $-E_{12}$ and $-E^{\prime}_{12}$ are superlinearly increasing functions of  $-\omega_{12}$, whereas $-E_{13}$ and $-E^{\prime}_{13}$ have a maximum at a finite $-\omega_{12}$. The decrease of $-E_{13}$ and $-E^{\prime}_{13}$  with $-\omega_{12}$ is a direct consequence of the relative phase suppression between the particular components. The Rabi energies for other values of $\sigma$ show qualitatively similar behavior. 

\subsection{Domain walls in the relative phases}
\label{subsc:domain_walls}

\begin{figure}
\includegraphics[width=0.826\columnwidth,keepaspectratio]{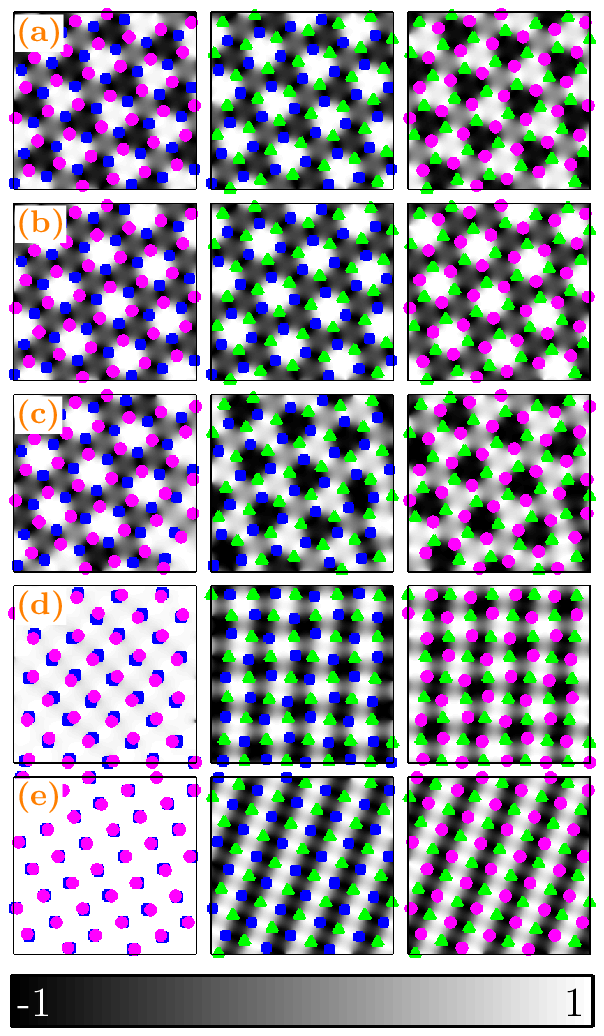}
\caption{\label{fig:domain} Relative phases between the three condensate components shown in terms of $\mathrm{sgn}\left(\omega_{ij}\right)\cos \phase_{ij}$, where $\phase_{ij}=\arg\left(\Psi_i\right)-\arg\left(\Psi_j\right)$ and the $\mathrm{sgn}$ function makes small values (shown in black) correspond to maximally repulsive pairwise Rabi energy density $\varepsilon_{ij}$. The first column is for $-\cos\phase_{12}$, the second for $\cos\phase_{13}$, and the third for $\cos\phase_{23}$. The rows correspond to different values of $\omega_{12}/\omega$: (a)~$-0.01$, (b)~$-0.03$, (c)~$-0.04$, (d)~$-0.06$, and (e)~$-0.12$. Other parameters are fixed at  $\sigma=0.7$, $\Omega/\omega =0.97$, $g = 2115$, and $\omega_{13}=\omega_{23}=0.01\omega$. The field of view in each panel is $11 a_r\times 11 a_r$, and the vortices in $\Psi_1$, $\Psi_2$, and $\Psi_3$ are marked with (blue) squares, (magenta) dots, and (green) triangles, respectively.}
\end{figure}

The Rabi coupling leads to well-defined relative phases between the condensates, and therefore, to the possibility of domain walls, i.e., one-dimensional defects~\cite{domain_wall_note}, in the relative phase fields~\cite{PhysRevA.65.063621}. The Rabi term breaks the U(1) symmetries associated with the relative phases $\phase_{ij}\equiv \phase_i-\phase_j$, where $\phase_i=\mathrm{arg}\left(\Psi_i\right)$ and $i<j$, by rendering the value for which $\mathrm{sgn}\left(\omega_{ij}\right) \cos \phase_{ij} =1$ energetically favorable. This prompts us to define a domain wall to lie along the path that connects two oppositely charged vortices in $\phase_{ij}$ (i.e., same-sign vortices in $\phase_i$ and $\phase_j$) and satisfies $\cos \phase_{ij} =-\mathrm{sgn}\left(\omega_{ij}\right)$, i.e., maximizes the phase-dependent part of the pairwise Rabi energy. Together, the repulsive density--density coupling $g_{ij}>0$ and the Rabi coupling $\omega_{ij}\neq 0$ give rise to an energy minimum at a finite domain-wall length~\cite{PhysRevLett.93.250406}. Increasing $|\omega_{ij}|$ decreases this optimal length until the two oppositely charged vortices  in $\phase_{ij}$ merge and the domain wall vanishes.

Let us now investigate the behavior of domain walls in the states discussed in the preceding sections~(Figs.~\ref{fig:s02_w12_001_set1}--\ref{fig:s1_w12_009_set1} and~\ref{fig:s02_w12_009_set2}--\ref{fig:s1_w12_009_set2}). To this end, we consider the three components  in pairs and their corresponding pairwise relative phases $\phase_{12}$, $\phase_{23}$, and $\phase_{13}$. The relative phases are presented in panels (g)--(i) of said figures using the quantities $\mathrm{sgn}\left(\omega_{ij}\right) \cos \phase_{ij}$, with domain walls shown in black; the positions of vortices of the relevant components are also indicated.

The properties of the domain walls depend on the strength of the Rabi coupling. For example, their characteristic width (analogous to the core size of vortices) in $\phase_{ij}$ is proportional to $|\omega_{ij}|^{-1/2}$~\cite{PhysRevA.65.063621}. In Fig.~\ref{fig:domain}, we show how the domain walls change when $\omega_{12}$ is varied in the range $0.01\omega \leq -\omega_{12} \leq 0.12 \omega$ while the other parameters are kept constant (so that the depicted states lie along the vertical line $\sigma=0.7$ in Fig.~\ref{fig:phase_diagram_set1}). At $\omega_{12}=-0.01\omega$, the domain walls are fairly delocalized, appearing wide between the oppositely charged vortices in each $\phase_{ij}$ [Fig.~\ref{fig:domain}(a)]. Increasing $|\omega_{12}|$ to $0.03\omega$ narrows the domain walls in $\phase_{12}$, while $\phase_{13}$ and $\phase_{23}$ remain nearly unchanged [Fig.~\ref{fig:domain}(b)]. In this regime, the pairwise Rabi energies $-E_{12}$, $-E_{13}$ and $-E_{23}$ all increase with increasing $-\omega_{12}$, as shown in Fig.~\ref{fig:rabi}. Figures~\ref{fig:domain}(a)--\ref{fig:domain}(c) all correspond to the lattice phase that consists of three mutually interlaced triangular lattices, has $\avrtopind=1$, and is illustrated in Fig.~\ref{fig:s02_w12_001_set1}. At $\omega_{12}=-0.06\omega$, the strong Rabi coupling between components~1 and~2 shrinks the domain walls in $\phase_{12}$, with vortices in $\phase_1$ and $\phase_2$ almost coinciding [Fig.~\ref{fig:domain}(d)]. Simultaneously, the Rabi energies $-E_{13}$ and $-E_{23}$ reach their maximum and gradually start decreasing due to the relative phase frustration occurring for these pairs (Fig.~\ref{fig:rabi}). The state in Fig.~\ref{fig:domain}(d) has $\avrtopind=4/3$ and a rectangular vortex-lattice unit cell enclosing two vortices of each component. Finally, at $\omega_{12}=-0.12\omega$ [Fig.~\ref{fig:domain}(e)], the coincidence of vortex positions between components~1 and~2 has become almost perfect and the domain walls have essentially vanished in $\phase_{12}$~\cite{wall_length_note}; this also halves the size of the elementary unit cell, yielding $\avrtopind=2/3$. Thus, we arrive at a peculiar state in which the domain walls persist in $\phase_{13}$ and $\phase_{23}$, but vanish completely in $\phase_{12}$. The effective locking of components~1 and~2 with $\phase_{12} \simeq \pi$ implies that $\phase_{13} \equiv \phase_{23}+\phase_{12} \simeq \phase_{23}+\pi$, in agreement with Fig.~\ref{fig:domain}(e).

\subsection{Lattice phases revisited}
\label{subsc:lattice_phases}

Equipped with the insight gained from the previous two subsections, let us return to the  phase diagrams in Figs.~\ref{fig:phase_diagram_set1} and~\ref{fig:phase_diagram_set2}, and the various skyrmionic phases therein. One can see that in the limit of large $|\omega_{12}|$, both phase diagrams exhibit a hexagonal-to-rectangular transition in the underlying vortex-lattice geometry, which is qualitatively similar to the transition observed in density--density-coupled two-component BECs~\cite{PhysRevLett.88.180403,PhysRevLett.91.150406}. In order to understand how it comes about in the three-component system, note that when $|\omega_{12}|$ is large enough to overcome the density--density repulsion due to $g_{12}>0$ and dominate over the other Rabi couplings, components~1 and~2 become effectively locked together such that $\Psi_1 = \mathrm{sgn}\left(\omega_{12}\right)\Psi_2=-\Psi_2$. At the same time, the Rabi coupling becomes very weak for the pairs~1--3 and~2--3 because of the suppression effect; in fact, since we have $\omega_{13}=\omega_{23}$, $\Psi_1 = -\Psi_2$ implies that $\varepsilon_{13}=-\varepsilon_{23}$, leading to cancellation of these Rabi couplings from the energy functional. As a consequence, in this fully locked limit components~1 and~2 can be viewed as a single component, and the system starts to behave like a repulsive two-component system with only density--density interactions. Then the hexagonal-to-rectangular transition is expected in the overall lattice geometry, and the ensuing ground states can be classified according to the results of Refs.~\cite{PhysRevLett.88.180403,PhysRevLett.91.150406}.
 
For the parameter set used in Fig.~\ref{fig:phase_diagram_set1}, an example from the two-component-like regime with triangular lattices is presented in Fig.~\ref{fig:s02_w12_009_set1} ($\sigma=0.2$), while Fig.~\ref{fig:domain}(e) shows a two-component-like state with square lattices ($\sigma=0.7$). The locking of components~1 and~2 implies that no additional lattice phases are expected for $|\omega_{12}| > 0.12\omega$ in Fig.~\ref{fig:phase_diagram_set1}. For the parameter set of Fig.~\ref{fig:phase_diagram_set2}, on the other hand, the Rabi frequencies $\omega_{13}=\omega_{23}=0.05\omega$ are so large that for the values of $\omega_{12}$ considered, the two-component limit with fully locked components~1 and~2 is reached only for $\sigma \leq 0.2$ (Fig.~\ref{fig:s01_w12_016_set2}). Nevertheless, increasing $|\omega_{12}|$ beyond the value $0.16$ shown in Fig.~\ref{fig:phase_diagram_set2} is expected to eventually result in two-component-like lattice phases with $\avrtopind=2/3$ also for $0.2 < \sigma <1$.  

The states containing zig-zag vortex lattices in some of the components (Figs.~\ref{fig:s02_w12_009_set2}, \ref{fig:s07_w12_003_set2}, and~\ref{fig:s07_w12_012_set2}) appeared in the regime where all $|\omega_{ij}|$ were comparable with each other and with the density--density repulsions. The zig-zag lattices can be viewed as deformed Abrikosov lattices where vortices originally in a straight row have been displaced in alternating directions. In the state shown in Fig.~\ref{fig:s02_w12_009_set2}, with $\sigma=0.2$ and $\avrtopind=4/3$, these displacements are in opposite directions in components~1 and~2. Visual inspection of Figs.~\ref{fig:s02_w12_009_set2}(g)--\ref{fig:s02_w12_009_set2}(i) reveals that the zig-zag configuration can efficiently accommodate relatively tightly bound vortex dimers in each $\phase_{ij}$, rendering it the energy-minimizing state for the comparable  Rabi  and density--density couplings between the components. It also follows from the zig-zag pattern that the dimers in each $\phase_{ij}$ are arranged in an antiferromagnetic order relative to one another, maximizing the intracomponent vortex distances. The states illustrated in Figs.~\ref{fig:s07_w12_003_set2} and~\ref{fig:s07_w12_012_set2} can be understood in a similar way but with the vortex lattices in the individual components having an underlying square geometry instead of the hexagonal one observed in Fig.~\ref{fig:s02_w12_009_set2}~\cite{unit_cell_note}. Zig-zag vortex patterns have previously been found for single-component BECs in highly eccentric harmonic trap potentials~\cite{PhysRevA.83.053612}.

\section{Conclusions}
\label{sc:conclusions}

In this work, we have shown that Rabi-coupled three-component BECs can host unconventional vortex lattices in the rotating ground state of the system. Such lattices were found to involve, for example, vortices arranged in square, zig-zag, or chain patterns, or coalesced into dimers or doubly quantized fused-core vortices. Based on the elementary unit cell of the combined lattice pattern in each state, we classified the ground states as either hexagonal or rectangular. We also argued that the emerging multicomponent lattices can be interpreted as having a skyrmionic character, which we quantified in terms of a topological index computed over the unit cell. By combining these two classification schemes, we collected our results in the phase diagrams of Figs.~\ref{fig:phase_diagram_set1} and~\ref{fig:phase_diagram_set2}. They enable one to identify the proper ground-state skyrmionic vortex lattice for different values of the intercomponent interaction strength $\sigma$ and the Rabi frequency $\omega_{12}$, for the two different parameter sets we have used in this work. 

For certain combinations of signs and values of the Rabi frequencies $\omega_{ij}$, some of the pairwise Rabi energies turned out to be heavily suppressed due to relative phase frustration. For example, when $\omega_{12}\omega_{13}\omega_{23} < 0$ and $|\omega_{12}|\gg |\omega_{13}|\approx |\omega_{23}|$, the Rabi coupling is significant only between components~1 and~2. Such Rabi suppression results in an effective reduction of the three-component BEC to a density--density-coupled two-component BEC. In this limit, the three-component system was found to exhibit a hexagonal-to-rectangular vortex-lattice transition in agreement with the previous results for the repulsively coupled two-component BECs and the classification given in Refs.~\cite{PhysRevLett.88.180403,PhysRevLett.91.150406}. Similarly, a three-component multiband superconductor, for which the Josephson-type coupling serves as an analog of the Rabi coupling in BECs, also reduces to a two-gap superconductor at a particular choice of the coupling matrix~\cite{PhysRevB.81.134522}.
 
Considering that we have limited our study to a specific subset of the large parameter space (e.g., by fixing $g$ and assuming $0 < \sigma \leq 1$), we expect the rotating ground state of the Rabi-coupled three-component BEC to harbor many more unforeseen vortex-lattice structures. This should be especially true in the more general case where one relaxes the equalities $g_{11}=g_{22}=g_{33}$ and $g_{12}=g_{13}=g_{23}$, which tend to favor equal populations of the three components.

In light of the Rabi suppression, it might also be interesting to compare and contrast the present system with a hybrid three-component system in which two components are different spin states of the same atom coherently coupled to each other, while the third component is a different species with different atomic mass and coupled only through density--density interactions to the other two components. In the limit of strong Rabi coupling, the hybrid system would reduce to a mass-imbalanced two-component  BEC, whose rotating ground state exhibits a variety of unconventional ground-state vortex structures~\cite{PhysRevA.85.043613,PhysRevA.91.043605}. This more intricate limiting behavior suggests that when the Rabi coupling is of intermediate strength, rotation of the hybrid system will likely produce novel ground-state vortex lattices that do not exist in the equal-mass system.

To experimentally realize our skyrmionic vortex lattices, one could use, e.g., ${}^{87}$Rb atoms in the ground-state multiplet $5^{2}S_{1/2}$. Two-component BECs have already been realized using mixtures of the $\ket{F=1,m_F=-1}$ and $\ket{2,1}$ sublevels~\cite{PhysRevLett.83.2498}, the $\ket{1,1}$ and $\ket{2,2}$ sublevels~\cite{PhysRevLett.106.065302}, and the $\ket{2,1}$ and $\ket{2,2}$ sublevels~\cite{PhysRevLett.85.2413} from this multiplet. The Rabi-coupled three component BEC could be created by optically trapping a mixture of three of these states and using  microwave and radiofrequency radiation to coherently couple them via two-photon processes. The vortex lattices could then be produced by inducing rotation in the system and letting it relax to its minimum-energy state. Measurement of the relative phase angles between the BEC components should  also be feasible~\cite{PhysRevLett.81.1543}. Furthermore, we note that many of the discovered skyrmionic phases could be distinguished by imaging only the profile of the total density $\ntot$.

\begin{acknowledgments}
This work was supported by the Research Foundation - Flanders (FWO). P.~K. acknowledges financial support from the Emil Aaltonen Foundation, the Finnish Cultural Foundation, the Magnus Ehrnrooth Foundation, and the Technology Industries of Finland Centennial Foundation. The authors thank R.~P.~Anderson, E.~Babaev, I.~O.~Cherednikov, V.~R.~Misko, T.~P.~Simula, and J.~Tempere for useful comments and discussions. 
\end{acknowledgments}
\bibliography{thc_bec_rabi_final.bbl}

\begin{thebibliography}{65}
\bibitem [{hex()}]{hexagonal_note}%
  \BibitemOpen
  \href@noop {} {}\bibinfo {note} {We call this lattice phase ``hexagonal"
  rather than ``triangular" to reflect its inclusion of unconventional states
  such as the one shown in Fig.~\ref{fig:s1_w12_009_set1}.}\BibitemShut {Stop}%
\bibitem [{Qij()}]{Qij_note}%
  \BibitemOpen
  \href@noop {} {}\bibinfo {note} {In other words, $\mathcal{Q}_{ij}$ is
  computed from Eq.~\eqref{eq:top_ind} with $\Psi_k=0$ for
  $k\notin\left\{i,j\right\}$.}\BibitemShut {Stop}%
\bibitem [{dom()}]{domain_wall_note}%
  \BibitemOpen
  \href@noop {} {}\bibinfo {note} {The domain walls are one dimensional (and
  vortices are zero dimensional) because our system is two
  dimensional.}\BibitemShut {Stop}%
\bibitem [{wal()}]{wall_length_note}%
  \BibitemOpen
  \href@noop {} {}\bibinfo {note} {The residual domain-wall lengths are smaller
  than or comparable to the uncertainty in the vortex positions and can thus be
  neglected.}\BibitemShut {Stop}%
\bibitem [{uni()}]{unit_cell_note}%
  \BibitemOpen
  \href@noop {} {}\bibinfo {note} {Note, however, that the unit cell of the
  \emph{combined} vortex lattice is rectangular for all three
  states.}\BibitemShut {Stop}%
\end{thebibliography}
\end{document}